\documentclass[9pt]{article}
\usepackage{spconf,amsmath,graphicx,hyperref}
\usepackage{spconf,amsmath,graphicx}
\usepackage{times}
\usepackage{epsfig}
\usepackage{amsmath}
\usepackage{amssymb}
\usepackage{algorithm}
\usepackage{algorithmic}
\usepackage{enumerate}
\usepackage{multirow}
\usepackage{multicol}
\usepackage{arydshln}
\usepackage{color}
\usepackage{amssymb}
\usepackage{soul}
\usepackage{subfig}
\usepackage{subcaption}
\usepackage{caption}
\usepackage{float}
\usepackage{booktabs}
\usepackage{babel,blindtext}
\usepackage{microtype}

\def\0{{\mathbf 0}}
\def\1{{\mathbf 1}}

\def\x{{\mathbf x}}
\def\y{{\mathbf y}}

\def\C{{\mathbf C}}
\def\D{{\mathbf D}}

\def\L{{\mathbf L}}

\def\U{{\mathbf U}}
\def\V{{\mathbf V}}
\def\W{{\mathbf W}}
\def\X{{\mathbf X}}
\def\Y{{\mathbf Y}}

\def\ie{{\textit{i.e.}}}

\def\cE{{\mathcal E}}

\def\cG{{\mathcal G}}

\def\cO{{\mathcal O}}

\def\bLambda{{\boldsymbol \Lambda}}

\newcommand{\up}{$\uparrow$}
\newcommand{\dn}{$\downarrow$}

\title{TSGL: Teacher-Student Graph Learning for 3DGS Compression}
\name{Matin Bani Saedi, Matthew Kyan, Gene Cheung}

\address{York University, Canada}
\begin{document}
\ninept
\maketitle
\suppressfloats[t]
\begin{abstract}
3D Gaussian Splatting (3DGS) is a popular representation for novel view synthesis. 
However, 3DGS contains millions of Gaussian primitives, each with rich attributes, resulting in large file sizes. 
We propose a novel 3DGS compression method based on Teacher-Student Graph Learning (TSGL) that operates directly on a trained model, without 3DGS retraining or access to training images.
Specifically, for each block of Gaussian primitives, using decoded positions and DC spherical harmonic (SH) coefficients as predictors, we learn a signal-dependent geometry graph $\cG$ encoding the pairwise similarities between neighbouring Gaussians via a teacher-student model.  
Given $\cG$, we perform Graph Fourier Transform (GFT) on the remaining attributes, so that signal energies are predominantly projected into the low-frequency coefficients for compact representation. 
On three standard benchmarks, the method reaches $27\times$ to $33\times$ compression with less than 0.6 dB of PSNR loss, improving on recent post-training compression methods in both size and rendering quality.
\end{abstract}

\begin{figure}[t]
    \centering
    \includegraphics[width=\linewidth]{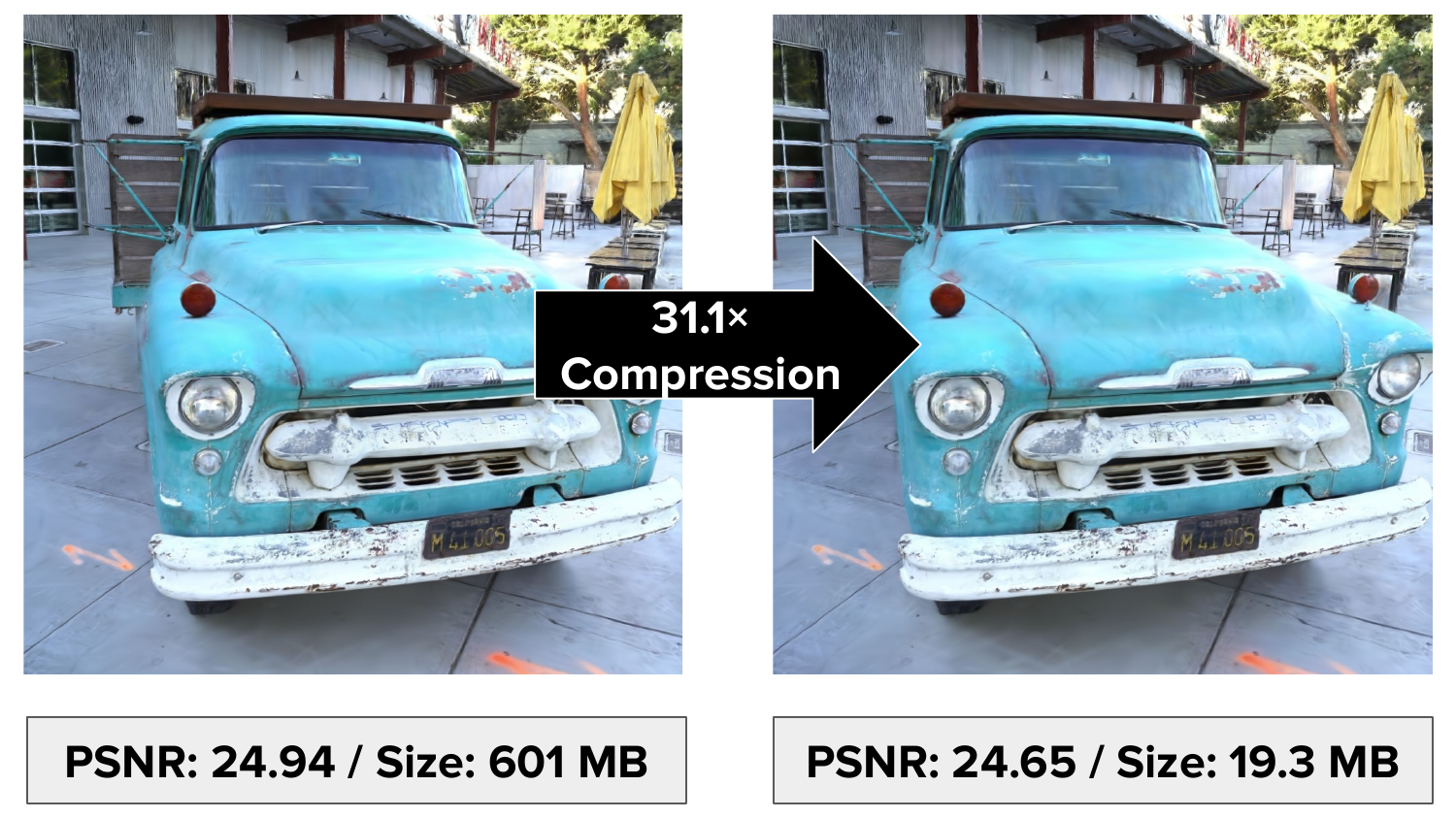}
    \caption{Qualitative comparison between the original 3DGS model (left) and our compressed model (right). Our method achieves $31.1\times$ compression with only a small decrease in PSNR.}
    \label{fig:qualitative}
\end{figure}

\section{Introduction}
\label{sec:intro}

Novel view synthesis reconstructs a scene from a set of reference images so that it can be viewed from any viewpoint, supporting applications in virtual reality, robotics, and digital twins. The field has progressed rapidly, moving from implicit neural representations such as Neural Radiance Fields (NeRF)~\cite{mildenhall2020nerf} to explicit primitive-based models such as 3D Gaussian Splatting (3DGS)~\cite{kerbl2023gaussians}. A 3DGS model consists of millions of Gaussian primitives, each defined by position,
rotation and scale vectors, an opacity scalar, and spherical harmonic (SH) coefficients that
encode view-dependent colour. Despite its rendering quality and speed, the large number of primitives makes 3DGS costly
to store and transmit. 

To tackle this challenge, methods either build a compact representation during
training~\cite{chen2024hac, lu2024scaffoldgs, lee2024compact, morgenstern2024sog} or
compress an already-trained model~\cite{fan2023lightgaussian, niedermayr2024compressed, xie2024mesongs, huang2026entropygs, tian2025flexgaussian}. 
In the first category, Scaffold-GS~\cite{lu2024scaffoldgs} learns anchor points with associated features and offsets, then uses small neural networks to predict local Gaussian attributes. HAC~\cite{chen2024hac} builds on Scaffold-GS by adding a learned spatial hash grid that provides contextual information for each anchor. 
Compact3DGS~\cite{lee2024compact} reduces storage through learnable Gaussian masks and a shared neural field for view-dependent colour. In the second category, LightGaussian~\cite{fan2023lightgaussian} combines
importance-based pruning with knowledge distillation to learn a lower-degree SH representation while preserving the original appearance. MesonGS~\cite{xie2024mesongs} employs the region-adaptive hierarchical
transform (RAHT) on selected Gaussian attributes for more efficient coding. EntropyGS~\cite{huang2026entropygs} uses statistical models of Gaussian attributes to design an entropy-coding scheme tailored to the
characteristics of 3DGS data.

Unlike previous works, we focus on  designing \textit{transform coding} for 3DGS attribute compression.
Standard transforms such as the \textit{discrete cosine transform} (DCT) used in JPEG~\cite{wallace1992jpeg} assume data are sampled on a regular 2D grid. 
In contrast, Gaussian primitives are irregularly distributed in 3D space, making the \textit{graph Fourier transform} (GFT) \cite{ortega18ieee}---where transformation is defined for discrete signals on an irregular kernel described by a graph---a suitable choice.  
GGSC~\cite{yang2024ggsc} introduces a graph-based compression baseline for 3DGS: i) partition the scene into local blocks, ii) construct a graph for each block using a Gaussian kernel based on primitive positions, and iii) apply GFT to Gaussian attributes for compression. 
L-GGSC~\cite{kuwabara2025lggsc} extends this baseline by introducing a parameterized graph shift operator, while retaining the same position-based graph. 
However, the effectiveness of GFT in energy compaction strongly depends on the underlying graph that encodes pairwise similarities / correlations. 
Edge weights based solely on spatial positions are clearly insufficient: close primitive neighbours can have attributes that differ greatly.

\begin{figure*}[!t]
    \centering
    \includegraphics[width=0.9\textwidth]{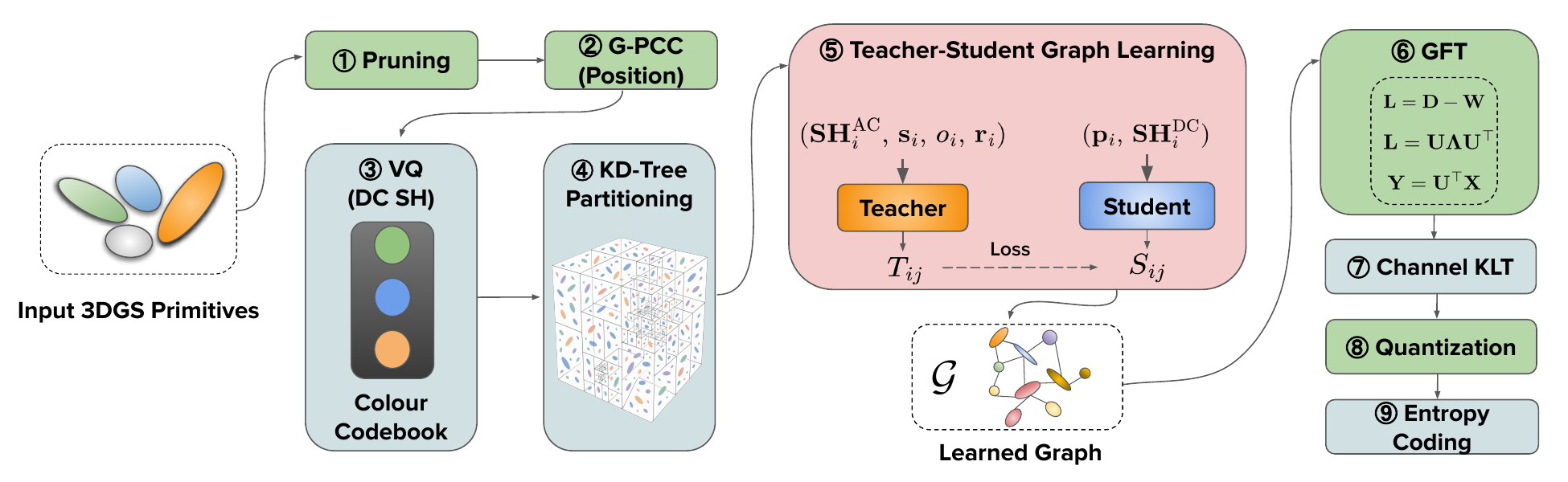}
    \vspace{-0.15in}
    \caption{Overview of the proposed 3DGS compression pipeline.}
    \label{fig:pipeline}
\end{figure*}

In response, we propose \textit{Teacher-Student Graph Learning} (TSGL) for 3DGS compression, which operates directly on a trained model without retraining or access to the training images. 
The key idea is the following:
\begin{quote}
Given a \textit{teacher} graph that encodes the ground-truth signal \textit{geometry}---\ie, a large (small) edge weight $w_{i,j}$ when target attribute difference $|x_i - x_j|$ is small (large)---we learn a \textit{student} graph via a lightweight network to mimic the teacher graph using only available information at the decoder (decoded Gaussian positions and DC SH coefficients) as predictors.     
\end{quote}
Using the constructed student graph, we apply the corresponding GFT to encode the remaining attributes for compression. 
Note that unlike \textit{Karhunen-Lo\`{e}ve Transform} (KLT) that optimally decorrelates signals \textit{statistically} in the aggregate~\cite{goyal2001transform}, our TSGL constructs one graph of signal geometry \textit{deterministically} for each individual signal $\x$ for low-frequency concentration and therefore compact representation. 
To our knowledge, this is the first method to \textit{directly} learn a graph of signal-dependent geometry for 3DGS compression rather than a generic similarity graph based on spatial locations. 
Experiments on 13 scenes from three standard benchmarks show $27\times$
to $33\times$ compression within 0.6 dB of the original models in PSNR, improving upon the rate-distortion performance of recent post-training methods.

\section{Preliminaries}
\label{sec:preliminaries}

\subsection{3D Gaussian Splatting}

Each 3DGS primitive $i$ is represented by
\begin{equation}
\mathbf{g}_i =
(\mathbf{p}_i, o_i, \mathbf{r}_i, \mathbf{s}_i, \mathbf{SH}_i),
\end{equation}
where $\mathbf{p}_i\in\mathbb{R}^3$ is the position, $o_i \in \mathbb{R}$ is the opacity,
$\mathbf{r}_i\in\mathbb{R}^4$ is the rotation quaternion,
$\mathbf{s}_i\in\mathbb{R}^3$ is the scale, and $\mathbf{SH}_i \in \mathbb{R}^{48}$ is the
spherical harmonic (SH) coefficient vector.
$\mathbf{SH}_i$ is composed of a DC component $\mathbf{SH}_i^{\mathrm{DC}}\in\mathbb{R}^3$, which represents the base colour of the Gaussian, and higher-order AC components
$\mathbf{SH}_i^{\mathrm{AC}}\in\mathbb{R}^{45}$, which capture colour variations with viewing direction.

\subsection{Graph Signal Processing}

Following notations in \cite{ortega18ieee}, an undirected weighted graph is defined as
$\mathcal{G}=(\mathcal{V},\mathcal{E},\W)$, where $\mathcal{V} = \{1, \ldots, N\}$ is the set of $N$ nodes, $\mathcal{E} = \{(i,j)\}$ is the set of edges, and $\W\in\mathbb{R}^{N\times N}$
is the \textit{adjacency} matrix. 
Each entry $w_{i,j}\geq 0$ represents the weight of an edge connecting nodes $i$ and $j$ if $\exists (i,j) \in \cE$, with $w_{i,j}=0$ if $\not\exists (i,j) \in \cE$.
$\W$ is symmetric, \ie, $w_{i,j}=w_{j,i}, \forall i,j$, since $\cG$ is undirected. 
A multi-channel graph signal is represented by $\X\in\mathbb{R}^{N\times c}$, where each column assigns one value to every node. 
The \textit{combinatorial graph Laplacian} matrix $\L \in \mathbb{R}^{N \times N}$ is defined as
$\L \triangleq \D - \W$, where $\D \in \mathbb{R}^{N \times N}$ is the diagonal \textit{degree} matrix with diagonal entries
$D_{i,i}=\sum_j w_{i,j}$.

\subsection{Graph Fourier Transform}

Since $\L$ is symmetric, by the Spectral Theorem \cite{Horn2012} it can be eigen-decomposed into
$\L = \U \bLambda \U^\top$, where the columns of $\U$ are the eigenvectors of $\L$, and
$\bLambda = \operatorname{diag}(\lambda_1,\ldots,\lambda_N)$
contains the corresponding eigenvalues in ascending order,
$\lambda_1 \leq \cdots \leq \lambda_N$. The eigenvectors form the graph Fourier basis, while the eigenvalues represent the associated graph frequencies. The GFT of a signal $\X$ and its inverse are then given by 
\begin{equation}
\Y = \U^\top \X, \qquad \X = \U \Y,
\label{eq:gft}
\end{equation}
where $\Y \in \mathbb{R}^{N\times c}$ contains the GFT coefficients.

\section{Proposed Coding Framework}
\label{sec:framework}

Our proposed coding framework is shown in Figure~\ref{fig:pipeline}. 
It begins with opacity- and importance-based pruning, followed by coding of the positions and DC SH coefficients, and spatial partitioning into local blocks. 
For each block, our proposed teacher-student graph learning method in
Section~\ref{sec:graph_learning} estimates the graph edge weights among the
primitives, and the learned graph defines the GFT basis. The GFT is applied to the remaining
attributes, namely the AC SH coefficients, scale, opacity, and rotation. 
A channel-wise KLT is then applied to the transformed AC SH coefficients to decorrelate inter-channel information, followed by quantization and entropy coding.

\subsection{Pruning, Base Attribute Coding, and Spatial Partitioning}

The pipeline begins with two pruning stages. 
We first remove Gaussian primitives whose
activated opacity values are below a fixed threshold of $0.01$. 
We then rank the remaining primitives by the importance score introduced in LightGaussian~\cite{fan2023lightgaussian} and discard $20\%$ of those with the lowest scores. 
This score accumulates each Gaussian's contribution to the rendered views and only requires the camera poses, not the
original training images or any parameter updates. 

Subsequently, the Gaussian positions are coded using G\nobreakdash-PCC~\cite{schwarz2019emerging} and the DC SH coefficients using $k$-means vector quantization with a scene-level codebook of $k=16384$ clusters. 
These two attributes are decoded at the decoder before graph construction and serve as inputs to the graph learning method in Section~\ref{sec:graph_learning}.

The pruned primitives are then partitioned into local blocks using a KD-tree built on the decoded positions, as in GGSC~\cite{yang2024ggsc}, so that the partition is reproduced at the decoder without requiring any side information. 
Since constructing the GFT basis requires eigen-decomposition of the graph Laplacian, which has a computational complexity of $\cO(N^3)$ for a block of $N$ primitives, operating on the entire scene would be computationally impractical. 
We therefore limit each KD-tree block to at most $200$ primitives.

\subsection{Transform Coding, Quantization, and Entropy Coding}

For each block, the graph learning method described in Section~\ref{sec:graph_learning} produces a weighted adjacency matrix $\W$. 
The corresponding graph Laplacian and GFT basis are then constructed as defined in Section~\ref{sec:preliminaries}.
The GFT is applied independently to the AC SH coefficients, scale,
opacity, and rotation along the primitive dimension.

The GFT exploits the relationships among Gaussian primitives but leaves the correlations across channels untouched. 
To decorrelate across channels, we compute a KLT basis from the covariance of the AC SH channels and apply it to the GFT coefficients. 
Let $M$ denote the total number of primitives in the scene and
$\overline{\mathbf{SH}}^{\mathrm{AC}}$ the mean of $\mathbf{SH}_i^{\mathrm{AC}}$
over the scene. The channel covariance is
\begin{equation}
\C = \frac{1}{M}\sum_{i=1}^{M}
  \bigl(\mathbf{SH}_i^{\mathrm{AC}}-\overline{\mathbf{SH}}^{\mathrm{AC}}\bigr)
  \bigl(\mathbf{SH}_i^{\mathrm{AC}}-\overline{\mathbf{SH}}^{\mathrm{AC}}\bigr)^{\top}
  \in \mathbb{R}^{45\times 45},
\label{eq:cov}
\end{equation}
whose eigenvectors, ordered by decreasing eigenvalue, form the KLT basis $\V$. 
The basis is computed once per scene and transmitted as \textit{side information}. 
Let
$\X_{\mathrm{ac}}\in\mathbb{R}^{N\times 45}$ denote the AC SH coefficients of a block.
Its transformed coefficients are
\begin{equation}
\Y_{\mathrm{ac}} = \U^{\top}\X_{\mathrm{ac}}\V ,
\label{eq:klt}
\end{equation}
where $\U$ operates along the primitive dimension while $\V$ operates along the channel dimension.

Finally, all transformed coefficients are uniformly quantized as
\begin{equation}
q = \operatorname{sign}(Y) \left\lfloor \frac{|Y|}{\Delta_a} + \theta_a \right\rfloor ,
\qquad
\hat{Y} = \Delta_a\, q ,
\label{eq:quant}
\end{equation}
where $q$ is the resulting integer index, $\Delta_a$ the quantization step for attribute $a$, $\theta_a$ the rounding offset, and $\hat{Y}$ the reconstructed coefficient. The quantized values are finally coded with a static rANS entropy coder~\cite{duda2013ans}, whose frequency tables are transmitted alongside the bitstream.

\section{Teacher-Student Graph Learning}
\label{sec:graph_learning}

\begin{table*}[t]
\centering
\footnotesize
\caption{Comparison with post-training 3DGS compression methods on the three standard
benchmarks. Best is shown in
\textbf{bold} and second best is \underline{underlined}. The 3DGS row is the uncompressed
reference and is not ranked.}
\label{tab:main}
\setlength{\tabcolsep}{3.5pt}
\begin{tabular*}{\textwidth}{@{\extracolsep{\fill}}l|cccc|cccc|cccc@{}}
\hline
& \multicolumn{4}{c|}{Mip-NeRF 360} & \multicolumn{4}{c|}{Tanks \& Temples} & \multicolumn{4}{c}{Deep Blending} \\
Method & PSNR\up & SSIM\up & LPIPS\dn & Size (MB)\dn & PSNR\up & SSIM\up & LPIPS\dn & Size (MB)\dn & PSNR\up & SSIM\up & LPIPS\dn & Size (MB)\dn \\
\hline
3DGS~\cite{kerbl2023gaussians} & 27.29 & 0.812 & 0.221 & 795.26 & 23.36 & 0.838 & 0.186 & 421.91 & 29.43 & 0.898 & 0.246 & 703.77 \\
\hline
GGSC~\cite{yang2024ggsc} & 18.42 & 0.449 & 0.498 & 71.21 & 16.43 & 0.480 & 0.489 & 29.29 & 25.67 & 0.807 & 0.386 & 109.85 \\
MesonGS~\cite{xie2024mesongs} & \underline{26.45} & \underline{0.791} & \underline{0.244} & 34.41 & \underline{22.95} & \textbf{0.825} & \textbf{0.202} & 17.56 & \textbf{29.13} & \textbf{0.894} & \textbf{0.255} & 29.52 \\
EntropyGS~\cite{huang2026entropygs} & 25.66 & 0.783 & 0.246 & \textbf{28.46} & 22.33 & \underline{0.806} & \underline{0.217} & \underline{15.23} & 27.48 & 0.882 & 0.266 & 25.95 \\
FlexGaussian~\cite{tian2025flexgaussian} & 26.38 & 0.780 & 0.251 & 40.80 & 22.44 & 0.804 & 0.219 & 16.30 & 28.61 & 0.884 & 0.269 & \underline{25.48} \\
\hline
\textbf{Ours} & \textbf{26.74} & \textbf{0.796} & \textbf{0.241} & \underline{29.32} & \textbf{23.09} & \textbf{0.825} & \textbf{0.202} & \textbf{14.25} & \underline{29.07} & \underline{0.892} & \underline{0.259} & \textbf{21.55} \\
\hline
\end{tabular*}
\end{table*}

The crux of our proposal is the \textit{signal-dependent} construction of the underlying graph $\cG$ on which the GFT is defined.
Unlike the KLT that decorrelates a signal $\x$ \textit{statistically} on the aggregate, our GFT---one constructed for individual $\x$---concentrates the energy of $\x$ in the low-frequency coefficients. 

Mathematically, low-frequency representation of $\x$ on graph $\cG$ with Laplacian $\L = \U \bLambda \U^\top$ means that only GFT coefficients $\y = \U^\top \x$ associated with small eigenvalues $\lambda_k$ are large in magnitude. 
In other words, projection of $\x$ into the eigen-space of $\L$ lies predominantly in the low-frequency subspace, resulting in a small \textit{Rayleigh quotient} $R_{\L}(\x)$ for a unit-norm signal $\x$:  
\begin{align}
R_{\L}(\x) \triangleq \x^\top \L \x = \sum_{(i,j) \in \cE} w_{i,j} (x_i - x_j)^2 = \sum_k \lambda_k y_k^2.
\label{eq:Rayleigh}
\end{align}
Looking at \eqref{eq:Rayleigh}, a graph $\cG$ that induces a low-frequency representation of $\x$ is one where edge weight $w_{i,j}$ is small (large) when $|x_i - x_j|$ is large (small).
Unlike the statistically derived KLT, such a graph $\cG$ encodes the deterministic \textit{geometry} of signal $\x$. 
The purpose of our teacher-student graph learning module is to estimate this signal geometry given limited available information at the decoder.

\subsection{Candidate Set}

The proposed graph is learned in two stages. 
A fixed design rule determines the edge set
$\cE$, and then the student network assigns weights to edges in $\cE$.
For each primitive $i$ in a block, we form a
candidate set $\mathcal{N}(i)$ as its $k=16$ nearest neighbours in
position plus its $k=16$ nearest neighbours in DC colour. 
Both position and DC colour are available at the decoder, so $\mathcal{N}(i)$ is
reconstructed there without side information.

\subsection{Teacher Graph}

Denote by $\x_i^{(a)}$ the vector of attribute
$a$ of primitive $i$, with $a$ ranging over the AC SH coefficients, scale, opacity,
and rotation. 
Since one graph per block serves all four attributes, we replace the scalar difference $|x_i - x_j|$ in~\eqref{eq:Rayleigh} with a distance over the attributes jointly,
\begin{equation}
d_{i,j} = \sum_a \frac{1}{\sigma_a^2}
  \bigl\lVert \x_i^{(a)} - \x_j^{(a)} \bigr\rVert^2 ,
\label{eq:dist}
\end{equation}
where $\sigma_a^2$ is the total variance of attribute $a$. 
This normalization allows attributes of different scales to contribute comparably. By~\eqref{eq:Rayleigh}, a graph
that induces a low-frequency representation of the attributes assigns a large (small) weight to an edge with small (large) $d_{i,j}$.

Following this criterion, we construct a \textit{teacher} graph over each block from the true attributes at the encoder, whose edge
weights are 
\begin{equation}
T_{i,j} = \frac{\exp(-d_{i,j}/\tau)}
              {\sum_{l\in\mathcal{N}(i)}\exp(-d_{i,l}/\tau)},
\qquad j \in \mathcal{N}(i),
\label{eq:teacher}
\end{equation}
where $\tau$ is a hyperparameter controlling how sharply the distribution
concentrates on the nearest candidates. The teacher thus provides an attribute-informed representation of the signal
geometry of the block, which serves as the target that the student learns to
estimate.

\subsection{Student Network}

While the teacher is constructed directly from~\eqref{eq:teacher} and requires no training, the \textit{student} is computed by a small network trained per scene to estimate the signal geometry captured by the teacher graph, using \textit{only} information available at the decoder. In other words, the student must infer signal geometry from position and DC colour alone.

Denote by $\mathbf{f}_i = (\mathbf{p}_i, \mathbf{SH}_i^{\mathrm{DC}})$ the combination of decoded position and DC colour of primitive $i$, both normalized over the
scene. 
Denote by $\mathcal{K}(i)$ its $16$ nearest neighbours in position, and
$\mathbf{e}_{i,j} = (\mathbf{f}_i,\, \mathbf{f}_j-\mathbf{f}_i)$ describes
neighbour $j$ relative to $i$. Inspired by local aggregation in
PointNet++~\cite{qi2017pointnetpp}, the student encodes each primitive from
its neighbourhood into an embedding
\begin{equation}
\mathbf{z}_i = \psi\Bigl(\mathbf{f}_i,\;
    \max_{j\in\mathcal{K}(i)}\phi(\mathbf{e}_{i,j}),\;
    \operatorname*{mean}_{j\in\mathcal{K}(i)}\phi(\mathbf{e}_{i,j})\Bigr),
\label{eq:encoder}
\end{equation}
where $\phi$ and $\psi$ are shared MLPs.

Given the embeddings, the student assigns a score to each candidate pair $(i, j)$ as:
\begin{equation}
r_{i,j} = \mathrm{MLP}\bigl(\mathbf{z}_i+\mathbf{z}_j,\;
                           |\mathbf{z}_i-\mathbf{z}_j|,\;
                           \mathbf{z}_i \odot \mathbf{z}_j\bigr),
\label{eq:score}
\end{equation}
where the inputs are constructed to be symmetric in $i$ and $j$ so that $r_{i,j}=r_{j,i}$.
The pairwise scores are then normalized over each primitive's candidate set to give
the student's distribution:
\begin{equation}
S_{i,j} = \frac{\exp(r_{i,j})}
              {\sum_{l\in\mathcal{N}(i)}\exp(r_{i,l})},
\qquad j \in \mathcal{N}(i).
\label{eq:student}
\end{equation}

Since the training objective is for the student to match the teacher, we minimize the
cross-entropy between the two distributions,
\begin{equation}
J = -\frac{1}{{M}}
    \sum_{i} \sum_{j \in \mathcal{N}(i)} T_{i,j}\log S_{i,j},
\label{eq:loss}
\end{equation}
where $M$ is the total number of primitives in the scene.

At decode time, the same student network produces $S$ from the decoded position and DC colour. Since the graph must be undirected, we average the two directions to obtain the adjacency matrix:

\begin{equation}
w_{i,j} = \frac{S_{i,j} + S_{j,i}}{2}.
\label{eq:adjacency}
\end{equation}
This adjacency matrix yields the graph Laplacian $\L$, whose eigenvectors give the GFT basis used in~\eqref{eq:gft}.
Since the student network is trained per scene, we transmit its parameters as side information.

\section{Experiments}
\label{sec:results}

\subsection{Experimental Settings}

We evaluate our method on 13 scenes from three datasets, Mip-NeRF 360~\cite{barron2022mipnerf360}, Tanks and Temples~\cite{Knapitsch2017}, and Deep Blending~\cite{hedman2018deepblending}, and report the average PSNR, SSIM, LPIPS and file size for each dataset. We compare against four recent post-training compression methods, GGSC~\cite{yang2024ggsc}, MesonGS~\cite{xie2024mesongs}, EntropyGS~\cite{huang2026entropygs}, and FlexGaussian~\cite{tian2025flexgaussian}, and include the uncompressed 3DGS models as a reference. GGSC, MesonGS, and EntropyGS are re-evaluated with our renderer on the same test views and at the same resolution, using their reported operating points, while the FlexGaussian numbers are quoted from the original paper. Since our method does not perform any 3DGS fine-tuning, we disable the corresponding re-optimization stages in the reproduced competing methods to ensure a fair comparison under the same post-training setting.

All sizes reported for our method correspond to the complete bitstream, including the G-PCC coded positions, the DC codebook and its per-primitive indices, the quantized and entropy-coded AC SH, scale, rotation and opacity coefficients, the student network weights, the KLT basis, and the rANS frequency tables. 

 The student network is trained independently for each scene for 20 epochs using Adam optimizer with a learning rate of $10^{-3}$ and a batch size of 32 blocks. The network contains 50{,}945 parameters, corresponding to 204 KB of side information, or approximately 0.7\% of the average compressed Mip-NeRF 360 bitstream. Finally, for attribute quantization, we use
$\Delta_a=(0.15,0.15,0.3,0.04)$
and $\theta_a=(0.35,0.5,0.5,0.5)$ for AC SH, scale,
opacity, and rotation, respectively.

\subsection{Experimental Results}

\label{ssec:comparison}
Table~\ref{tab:main} reports the compression performance of our method and the competing techniques, averaged separately over the scenes of each dataset. Our method
achieves the best rendering quality among the compressed methods on Mip-NeRF 360 and Tanks
and Temples, with the highest PSNR and SSIM and the lowest LPIPS. On Mip-NeRF 360, our
average bitstream size is 29.32 MB compared with 34.41 MB for MesonGS, while improving PSNR
by 0.29 dB. On Tanks and Temples, our method matches or improves all three quality metrics while also producing
the smallest bitstream among all compared methods, 6.4\% below EntropyGS, the second
smallest. On Deep Blending, MesonGS achieves a
slightly higher PSNR by 0.06 dB, but requires 7.97 MB more storage on average. Our method
also maintains higher quality than GGSC, EntropyGS, and FlexGaussian across all three datasets. On Mip-NeRF 360, for
instance, EntropyGS is only 0.86 MB smaller on average, but has 1.08 dB lower PSNR.
Overall, our compressed models are approximately $27\times$ to $33\times$ smaller than the
original 3DGS representations, while the average PSNR degradation remains below 0.6 dB on
all three datasets.

\vspace{0.05in}
\noindent 
\textbf{Ablation Studies:}
Table~\ref{tab:ablation} presents a cumulative ablation of our compression pipeline across
the datasets, where each row adds one stage to the configuration above it. The results
show that opacity and importance pruning, G-PCC, and vector quantization of the DC SH
coefficients have negligible impact on visual quality, while reducing the original model
size by approximately 40-45\%. To evaluate the contribution of the graph-based GFT stage,
we include a baseline in which the remaining attributes are quantized and entropy coded
directly, without graph construction or transformation. Adding the learned-graph GFT
consistently improves both rate and quality. The average file size decreases from 33.07 to
30.43 MB, 15.95 to 14.57 MB, and 22.35 to 21.13 MB on Mip-NeRF 360, Tanks and Temples, and
Deep Blending, respectively, while PSNR improves by 0.69, 0.46 and 0.78 dB. The channel KLT
improves PSNR by a further 0.35, 0.19 and 0.16 dB. It also reduces the bitstream by 3.6\% and 2.2\% on the first
two datasets, while on Deep Blending it costs 2.0\% in size.

\begin{table}[t]
\centering
\footnotesize
\setlength{\tabcolsep}{1.8pt}
\caption{Contribution of each pipeline stage, per dataset. Each row adds one
stage to the row above. Sizes are in MB.}
\label{tab:ablation}
\begin{tabular}{l|cc|cc|cc}
\hline
\multirow{2}{*}{Stage}
 & \multicolumn{2}{c|}{Mip-NeRF 360}
 & \multicolumn{2}{c|}{T\&T}
 & \multicolumn{2}{c}{Deep Blending} \\
\cline{2-7}
 & PSNR\up & Size\dn & PSNR\up & Size\dn & PSNR\up & Size\dn \\
\hline
Original 3DGS~\cite{kerbl2023gaussians} & 27.29 & 795.26 & 23.36 & 421.91 & 29.43 & 703.77 \\
+ Opacity pruning & 27.29 & 679.49 & 23.35 & 333.49 & 29.43 & 599.18 \\
+ Importance pruning & 27.28 & 543.59 & 23.35 & 266.79 & 29.43 & 479.34 \\
+ G-PCC, VQ DC & 27.27 & 476.37 & 23.34 & 234.02 & 29.41 & 419.51 \\
+ Coding (no transform) & 25.70 & 33.07 & 22.44 & 15.95 & 28.13 & 22.35 \\
+ GFT (learned graph) & 26.39 & 30.43 & 22.90 & 14.57 & 28.91 & 21.13 \\
+ Channel KLT & 26.74 & 29.32 & 23.09 & 14.25 & 29.07 & 21.55 \\
\hline
\end{tabular}
\end{table}

\section{Conclusion}
\label{sec:conclusion}
We proposed a novel compression technique for 3DGS based on Teacher-Student Graph Learning (TSGL). Our method trains a small student network to reproduce a teacher graph built from the true attributes using only decoder-available inputs, so that the GFT basis reflects the signal geometry without transmitting the graph itself. A scene-level channel-wise KLT is further applied to the AC SH coefficients, removing the correlation across channels. Experiments reveal that our proposed method outperforms recent post-training compression methods on most quality and size measurements across three
standard benchmarks.

\begin{small}
\bibliographystyle{IEEEbib}
\bibliography{refs}
\end{small}

\end{document}